\documentclass[10pt]{article}

\usepackage{%color,
graphicx,amsmath,amssymb,bm}

\newtheorem{prop}{Prop}[section]

\newtheorem{lemma}[prop]{Lemma}

\newtheorem{theorem}[prop]{Theorem}

\usepackage{amsmath,amssymb}
\usepackage{makeidx}
\makeindex
\usepackage{ascmac}
\usepackage{pb-diagram}
\usepackage{here}
\usepackage{esint}

\title{Existence of de Almeida-Thouless-type  
instability \\ 
 in the transverse field Sherrington-Kirkpatrick model
 }
 \author{C. Itoi $^1$, K. Fujiwara $^1$  and Y. Sakamoto$^2$\\
\\
Department of Physics,   
GS $\&$ CST, 
Nihon University
\\ $^2$Laboratory of Physics, CST, Nihon University
}
\begin{document}
\maketitle

\abstract{
 The interpolation method for mean field spin glass models
developed by Guerra and Talagrand 
is extended to a quantum mean field spin glass model.
This extension enables us to obtain both replica-symmetric (RS) and 
one step replica-symmetry breaking (1RSB) solutions of the free energy 
density in the transverse field Sherrington-Kirkpatrick model. 
It is shown that the RS %replica-symmetric 
solution is exact in the paramagnetic phase.
We provide a sufficient condition on 
coupling constants where the 1RSB % one step replica-symmetry breaking 
solution gives better bound than the RS %replica-symmetric 
one. 
This condition reduced to physical quantities in disordered single spin systems
allows a simple computer-assisted proof for the existence of 
the de Almeida-Thouless-type instability% in the transverse field Sherrington-Kirkpatrick model
. 
}
%\keywords{spin glass, replica symmetry breaking, overlap operator, mean field model, quantum spin}
\maketitle
%\setcounter{footnote}{0}
%%%%%%%% the end of preprint version
\tableofcontents%\maketitle
\section{Introduction} 
The transverse field Sherrington-Kirkpatrick (SK) model is well-known as one of the 
simplest quantum spin glass models, and has been studied extensively.
Several studies have been conducted  in mathematically rigorous methods \cite{AB,Cr,IISS,IS,W,LRRS}.
Recently, Leschke, Manai, Ruder and Warzel have proven that  
the variance of the overlap operator does not vanish in the transverse SK model \cite{W}
using the Falk-Bruch inequality \cite{FB,R} and the $\mathbb Z_2$-symmetry of the model. 
Their rigorous and striking result has been appreciated by many researchers studying spin glasses and quantum complex systems, since the finite variance of the overlap operator is recognized as a necessary condition for the existence of replica-symmetry breaking (RSB) \cite{IMT}.  
This has brought further attention to the interesting question of
whether the distribution of the overlap operator is broadened around the one of two peaks of the $\mathbb Z_2$-symmetric pair, 
since their argument relies  
on the fact that the expectation of the overlap operator vanishes due to the $\mathbb Z_2$-symmetry.

It is well-known that the square root interpolation method developed by 
Guerra and Talagrand is useful to obtain rigorous bounds on many physical quantities in the spin glass models \cite{G1,T}. 
This method gives the replica-symmetric (RS) and the RSB bounds on
the free energy density in SK model, rigorously.  
 In particular, one step RSB (1RSB) solution gives    
the de Almeida-Thouless (AT) line which is a phase boundary of the unstable region of the RS solution \cite{AT,T}.
To extend this method to quantum systems is interesting to study.

In the present paper, we obtain variational solutions of the free energy density in the 
 transverse field SK model. 
 We extend the square root interpolation method for RS and
 1RSB variational solutions of the
free energy density  given by Guerra and Talagrand \cite{G1,T}
to quantum mechanically perturbed models. First, we prove that the obtained RS solution 
becomes exact in the paramagnetic phase assuming the unbroken replica- and $\mathbb Z_2$-symmetries. 
For sufficiently low temperature and sufficiently weak transverse field, however, the finite variance of the
overlap operator \cite{W} enables us to prove that this paramagnetic RS solution cannot be exact.
In this case, our interest is possibility that another spin glass RS solution becomes exact.
Next, we construct a 1RSB solution, and find a condition on the unstable region of the RS solution,
where the 1RSB solution gives better bound on the free energy density than RS solutions.
If the condition is satisfied, the AT-type instability exists in the transverse field SK model. 
We represent a sufficient condition for 
AT-type instability in terms of disordered single spin systems, using the Falk-Bruch inequality \cite{FB,R}.
Then, a computer-assisted proof
by simple numerical calculations becomes possible to confirm this condition. 
This unstable region specified in the coupling constant space must be contained in the RSB phase.

The present paper is organized as follows. In section 2, we define the Hamiltonian and other physical quantities in the transverse field SK model. In section 3, the RS solution of the free energy density
in the transverse field SK model is obtained by the square root interpolation method extended to
quantum spin glass systems. The exactness and inexactness of the paramagnetic RS solution 
are shown even in this  quantum model in the paramagnetic phase, as in the classical SK model.
In section 4, the 1RSB solution of the free energy density in the transverse field SK model is obtained. 
In section 5, we obtain a sufficient condition that
the 1RSB solution gives better bound on the free energy density than the RS solution. 
This condition is confirmed numerically at several points in the coupling constant space.

\section{Definitions of the model}
Here, we study  quantum spin systems with random interactions. Let $N$ be a positive integer and a site index 
$i \ (\leq  N)$ is also a positive integer. 
A sequence of spin operators 
 $(\sigma^{w}_i)_{w=x,y,z, 1\leq i \leq  N}$ on a Hilbert space ${\cal H} :=\bigotimes_{i =1}^N {\cal H}_i$ is
defined by a tensor product of the Pauli matrix $\sigma^w$ acting on ${\cal H}_i \cong {\mathbb C}^{2}$ and unities.
These operators are self-adjoint and satisfy the commutation relation
$$
 [\sigma_k^y,\sigma_j^z]=2i \delta_{k,j} \sigma_j^x ,\ \ \  \ \ 
[\sigma_k^z,\sigma_j^x]=2i \delta_{k,j} \sigma_j^y ,\ \  \ \ \ [\sigma_k^x,\sigma_j^y]=2i \delta_{k,j} \sigma_j^z ,  
$$
and each spin operator satisfies
$$
(\sigma_j^w)^2 = {\bf 1}.
$$
The Sherrington-Kirkpatrick (SK) model is
  well-known as a disordered classical  spin  system \cite{SK}.
The transverse field  SK  model 
 is a simple quantum  extension.
Here, we study a magnetization process for a local field in these models.
Consider  the following Hamiltonian with  coupling constants
$b, c \in {\mathbb R}$, $c \geq 0$ 
\begin{equation}
H( \sigma, b, g):=- \frac{1 }{\sqrt{N}}\sum_{1\leq i<j\leq N} g_{i,j}  \sigma_i^z \sigma_j^z
-\sum_{j=1}^N b \sigma_j^x,
\label{hamil}
\end{equation}
where  $g=(g_{i,j})_{1\leq i<j \leq N}$are independent identically distributed (i.i.d) standard Gaussian random variables
obeying a probability density function
\begin{equation}
p(g):=  \prod_{1\leq i<j\leq N}
 \frac{1}{\sqrt{2\pi}} e^{-\frac{g_{i,j} ^2}{2}}
\label{distributiong}
\end{equation}
The Hamiltonian is invariant under $\mathbb Z_2$-symmetry $U \sigma_i ^z U^\dag = -\sigma_i^z$
for the discrete unitary transformation $U:= \prod_{1\leq i \leq N} \sigma_i^x$
.  
For a positive $\beta $, the  partition function is defined by
\begin{equation}
Z_N(\beta, b, g) := {\rm Tr} e^{ - \beta H(\sigma, b, g)},
\end{equation}
where the trace is taken over the Hilbert space ${\cal H}$.

\section{RS bound on the free energy density}
Guerra and Talagrand have provided the well-known square root interpolation method, which 
represents a variational solution of the free energy density in the classical mean field model
 in terms of that in the single spin model with suitable corrections \cite{G1,T}.
Here, we apply this method to the transverse field SK model, as for the SK model. 
Let  $(z_j)_{1\leq j \leq N} $ be a sequence of i.i.d standard Gaussian random variables. 
Consider the following interpolated  Hamiltonian  with  parameters $s \in [0,1]$ for $q \in [0,1]$
\begin{eqnarray}
H(s, \sigma):=
- \sqrt{\frac{s}{N}}\sum_{1\leq i<j\leq N} g_{i,j}  \sigma_i^z \sigma_j^z-\sum_{j=1}^N[ \sqrt{q(1-s)} z_j\sigma_j^z + b \sigma_j^x]. \label{hamils}
\end{eqnarray}
This interpolated Hamiltonian for $b=0$ is identical to that in the SK model obtained by Guerra and Talagrand \cite{G1,T}. 
Define an interpolated function $\varphi_N(s)$
\begin{equation}
\varphi_N(s) :=\frac{1}{N} \mathbb E \log {\rm Tr} e^{-\beta H(s, \sigma)}
\end{equation}
where $\mathbb E$ denotes the expectation over all Gaussian random variables $(g_{i,j})_{1 \leq i<j \leq N}$  and $(z_i)_{1\leq i\leq N}$.
Since the function $\varphi_N(1)$ is given by
\begin{equation}
\varphi_N(1) = \frac{1}{N}\mathbb E \log Z_N(\beta, b,g), \label{varphi1}
\end{equation}
the free energy density of the transverse field SK model is $-\varphi_N(1)/\beta$.
Let  $f$ be an arbitrary function 
of a sequence of spin operators $\sigma=(\sigma_i^w)_{ w=x,y,z, 1 \leq i \leq N}$.  
The  expectation of $f$ in the Gibbs state is given by
\begin{equation}
\langle f( \sigma) \rangle_s=\frac{{\rm Tr} f( \sigma)  e^{ - \beta H(s, \sigma)}}{{\rm Tr}  e^{ - \beta H(s, \sigma)}}.
\end{equation}
The derivative of $\varphi_N(s)$ with respect to $s$
is given by
\begin{equation}
\varphi'_N(s)= \frac{\beta}{2N^\frac{3}{2}\sqrt{s}} \sum_{1\leq < j\leq N} \mathbb E g_{i,j}\langle \sigma_i^z \sigma_j^z \rangle_s - \frac{\beta\sqrt{q}}{2N\sqrt{1-s}} \sum_{i=1}^N \mathbb E z_i \langle \sigma_i^z \rangle_s.
\end{equation}
Identities for the Gaussian random variables $g_{i,j}$ and $z_i$ and their probability distribution function
$$
g_{i,j} p(g, z) = -\frac{\partial p}{\partial g_{i,j}},    \ \ \ z_i p(g, z) =- \frac{\partial p}{\partial z_i}
$$
and the integration by parts imply
\begin{eqnarray}
\varphi'_N(s)&=& \frac{\beta^2}{2N^2} \sum_{1<i\leq < j\leq N} \mathbb E [(\sigma_i^z \sigma_j^z , \sigma_i^z \sigma_j^z )_s-\langle \sigma_i^z \sigma_j^z \rangle_s^2 ]- \frac{\beta^2q}{2N} \sum_{i=1}^N \mathbb E [(\sigma_i^z  , \sigma_i^z )_s-\langle \sigma_i^z \rangle_s^2 ] \nonumber \\
&=&\frac{\beta^2(N-1)}{4N}\mathbb E(\sigma_i^z \sigma_j^z , \sigma_i^z \sigma_j^z )_s
- \frac{\beta^2q}{2} \mathbb E (\sigma_i^z  , \sigma_i^z )_s -\frac{\beta^2}{4} \mathbb E \langle (R_{1,2}-q) ^2\rangle_s
+\frac{\beta^2}{4} \Big(q^2 +\frac{1}{N}\Big)
, \label{phi'}
\end{eqnarray}
where The Duhamel function for bounded linear operators $A, B$ is defined by
\begin{equation}
 (A,B) =\int_0^1 dt \langle e^{\beta t H} A e^{-\beta tH} B \rangle,
 \end{equation}
and  the overlap operator $R_{a,b}$ is defined by 
\begin{equation}
R_{a,b} := \frac{1}{N} \sum_{i=1}^N \sigma_i^{z,a} \sigma_i^{z,b},\label{R}
\end{equation}
for independent replicated Pauli operators $\sigma_i^{z,a} \ (a= 1, 2,\cdots, n)$ obeying the same Gibbs state with
the replica Hamiltonian
$$
H(s, \sigma^1, \cdots, \sigma^n):= \sum_{a=1}^n H(s, \sigma^a).
$$
This Hamiltonian is invariant under 
permutation of replica spins. This permutation symmetry is known to be the replica symmetry. 
The order operator $R_{a,b}$ measures the replica symmetry breaking as an order operator.  
Define a function
\begin{equation}
\rho(s, q):= \frac{(N-1)}{N}[1-\mathbb E(\sigma_i^z \sigma_j^z , \sigma_i^z \sigma_j^z )_s]
+ 2q [\mathbb E (\sigma_i^z  , \sigma_i^z )_s -1]
,
\end{equation}
which is non-negative valued.
The identity (\ref{phi'})  imply
\begin{eqnarray}
\varphi'_N(s) &=& \frac{\beta^2}{4}(1-q)^2-\frac{\beta^2}{4} \mathbb E \langle (R_{1,2}-q) ^2\rangle_s
-\frac{\beta^2}{4}\rho(s,q) \label{phi'2}
\end{eqnarray} 
Integration of this identity over $s\in[0,1]$ gives the following lemma.
{\lemma \label{L3.1} (Extended Guerra's identity for RS bound) \\
Define a function by
\begin{equation}
 \Phi(\beta,b,q):=\mathbb E \log 2 \cosh X(z,q) +\frac{\beta^2}{4}(1-q)^2-  \frac{\beta^2}{4} \int_0 ^1 ds\rho(s,q),\label{Phi}
\end{equation}
where the above random variable is defined by 
\begin{equation}
X(z,q):=\beta \sqrt{qz^2+ b^2}.  \label{X}
\end{equation}
For arbitrary $(\beta,b,q) \in [0,\infty)^2 \times [0,1]$,
the following identity is valid
\begin{eqnarray}
\varphi_N(1) 
 =\Phi(\beta,b,q)-\frac{\beta^2}{4} \int_0 ^1 ds 
\mathbb E \langle (R_{1,2}-q) ^2\rangle_s, \label{eq}
\end{eqnarray}
Proof.} Integration of the identity (\ref{phi'2}) over $s\in[0,1]$ gives
$$
\varphi_N(1) = \varphi_N(0) +\frac{\beta^2}{4} 
\int_0 ^1 ds  [(1-q)^2 -\rho(s,q) -\mathbb E \langle (R_{1,2}-q) ^2\rangle_s].
$$
The model at $s=0$ becomes independent spin model, and therefore $\varphi_N(0)$ is represented in terms of 
the partition function of a disordered single spin system
\begin{equation}
\varphi_N(0) = \mathbb E \log {\rm Tr} \exp \beta[ \sqrt{q}z \sigma^z + b \sigma^x]
=  \mathbb E \log 2 \cosh X(z,q).
\end{equation}
This completes the proof. $\Box$

Note that $\Phi(\beta,b,q)$ gives the following bound 
\begin{eqnarray}
\varphi_N(1) \leq \Phi(\beta,b,q),\label{Phiineq}
\end{eqnarray}
where the right hand side is called the RS bound.

To obtain lower and upper bounds on
$\Phi(\beta,b,q)$, let us evaluate $\rho(s,q)$.
The Falk-Bruch inequality \cite{FB,R} and a well-known inequality \cite{BT,S} for the Duhamel function
of an arbitrary bounded linear operator $A$ give
\begin{equation}
F\Big(\frac{\langle [A^\dag,[\beta H,A]\rangle_s}{2\langle \{A^\dag,A\}\rangle_s}\Big)\leq 
\frac{2(A^\dag,A)_s}{\langle \{A^\dag,A\}\rangle_s} \leq 1, 
\label{FB}
\end{equation}
where the function $F:[0, \infty) \to (0,1]$ is defined by 
\begin{equation}F(x \tanh x) = \frac{\tanh x }{x}, \label{FBF}
\end{equation}
and $F(0)=1$.
This function is  monotonically decreasing and convex.   
Therefore
\begin{eqnarray}
&& F(2\beta b\tanh \beta b)\leq F(\beta b (\langle \sigma_i^x\rangle_s+\langle \sigma_j^x\rangle_s)) =
F\Big(\frac{\beta}{4}\langle [\sigma_i^z\sigma_j^z,[H,\sigma_i^z\sigma_j^z]\rangle_s\Big)\leq(\sigma_i^z \sigma_j^z , \sigma_i^z \sigma_j^z )_s \leq 1, \label{D1} \\
&& \frac{\tanh \beta b}{\beta b}=F(\beta b\tanh \beta b) \leq  F(\beta b \langle \sigma_i^x\rangle_s) =F\Big(\frac{\beta}{4}\langle [\sigma_i^z,[H,\sigma_i^z]\rangle_s\Big)\leq
 (\sigma_i^z  , \sigma_i^z )_s\leq 1,\label{D2}
\end{eqnarray}
where an upper bound $\tanh \beta b \geq \langle \sigma_i^x \rangle_s$ has been used as shown by Leschke, Manai, Ruder and Warzel \cite{W}.
These inequalities (\ref{D1}), (\ref{D2}) and a well-known inequality given by
Dyson, Lieb and Simon \cite{DLS} 
\begin{equation}
F(t) \geq t^{-1}(1-e^{-t}), \label{Fineq}
\end{equation}
yield the following lower and upper bounds on the function $\rho(s,q)$
\begin{equation}
 2q\Big( \frac{\tanh \beta b}{\beta b}-1\Big)\leq \rho(s,q) \leq \frac{N-1}{N}
  \Big(1-\frac{1-e^{-2\beta b\tanh \beta b}}{2\beta b\tanh \beta b}\Big).
  \label{rhobounds}
\end{equation}
Lower and upper bounds (\ref{rhobounds}) on $\rho(s,q)$  give the following lemma for the RS bound.
 {\lemma \label{L3.2} The RS bound $\Phi(\beta,b,q)$ satisfies
 \begin{equation}
\Phi_L(\beta,b,q) \leq \Phi(\beta,b,q)\leq \Phi_U(\beta,b,q),
 \end{equation}
where lower and upper bounds are defined by
\begin{eqnarray}
&&\Phi_L(\beta,b,q):= \mathbb E \log 2 \cosh X(z,q)+\frac{\beta^2}{4}\Big[(1-q)^2-\Big( 1 -\frac{1-e^{-2\beta b\tanh \beta b}}{2\beta b\tanh \beta b}\Big)\Big]
   \label{lPhi}
 \\
  && \Phi_U(\beta,b,q):=\mathbb E \log 2 \cosh X(z,q) +\frac{\beta^2}{4} \Big[(1-q)^2 +2q\Big(1-\frac{\tanh \beta b}{\beta b}\Big)\Big].
\label{uPhi}
\end{eqnarray}}
A variational solution with the best bound
is obtained by minimizing the right hand side of 
(\ref{Phi}). The minimizer $q$ should satisfy
\begin{eqnarray}
0= \frac{\partial}{\partial q}\Phi(\beta, b,q)=\frac{\beta^2}{4}\Big[2\mathbb E\frac{ z^2 \tanh X(z,q)}{X(z,q)} 
+2q-2-\frac{\partial}{\partial q}
\int_0 ^1 ds \rho(s,q)
 \Big]. 
\label{Phi_q}
\end{eqnarray}
This minimizer $q$ gives the best bound on $\varphi_N(1)$ as a variational solution
\begin{eqnarray}
\varphi_N(1) &\leq& \min_{q\in[0,1]} \Phi(\beta,b,q) \nonumber \\
&=&\min_{q\in[0,1]} \Big[
\mathbb E \log 2 \cosh X(z,q) +\frac{\beta^2}{4}(1-q)^2-
 \frac{\beta^2}{4} \int_0 ^1 ds\rho(s, q) \Big],
\label{RS}
\end{eqnarray}
$\min_{q\in[0,1]} \Phi(\beta,b,q) $ is called RS solution. \\

Here, we discuss the exactness of the RS solution.
Assume that the replica-symmetry is unbroken and there exists $q \in [0,1]$
such that 
\begin{equation} 
\lim_{N\to \infty}\int_0^1ds\mathbb E \langle (R_{1,2}-q) ^2\rangle_s=0.
\label{R-q}
\end{equation}
The bound $\Phi(\beta, b,q)$ with the above $q$
gives the exact solution of $\varphi_\infty(1)$, as in the classical SK model. Only the minimizer $q$ of $\Phi(\beta, b,q)$
can give the equality,
since the inequality (\ref{Phiineq}) is valid for any $q$. In fact, this exactness can be shown in the case for $q=0$.
Consider the model for $q=0$ in the paramagnetic phase where the replica-symmetry and $\mathbb Z_2$-symmetry are unbroken. Substitute $q=0$ into the equation (\ref{eq}) in Lemma \ref{L3.1}, and the extended Guerra's identity becomes
\begin{eqnarray}
\varphi_N(1) &=&
\Phi(\beta,b,0) -\int_0^1ds \mathbb E \langle R_{1,2}^2\rangle_s  \label{RSparaeq}\\
&\leq&  \Phi(\beta,b,0)
=\log 2\cosh \beta b+ \frac{\beta^2}{4} -\frac{\beta^2}{4} \int_0 ^1 ds\rho(s,0).\label{RSpara}
\end{eqnarray}
Then, the following theorem is obtained. 
{\theorem \label{T3.3}(Exactness of the paramagnetic RS solution)\\
In the paramagnetic phase, $\Phi(\beta,b,0)$ gives the exact solution
\begin{equation}
\lim_{N\to\infty}\Phi(\beta,b,0)=\lim_{N\to\infty}\varphi_N(1).
\end{equation}
Proof.}
The existence of the right hand side in the infinite-volume limit is proven by \cite{AB,Cr}.  
The $\mathbb Z_2$-symmetry and the replica-symmetry imply
$$\langle R_{1,2}\rangle_s=0,  \ \ \ 
\displaystyle \lim_{N\to \infty}\mathbb E\langle (R_{1,2}-\mathbb E \langle R_{1,2}\rangle_s)^2\rangle_s=0
. 
$$  
These and the above identity (\ref{RSparaeq}) conclude that the paramagnetic RS solution is exact
$$\displaystyle\lim_{N\to\infty}\Phi(\beta,b,0)=\varphi_\infty(1).$$ 
This completes the proof. $\Box$ \\

Theorem \ref{T3.3} is consistent with the result in \cite{LRRS}. 
On the other hand, using the Falk-Bruch inequality (\ref{FB}), and assuming 
the ground state energy density $-\kappa \simeq -0.763$ of the classical SK model
and the $\mathbb Z_2$-symmetry $\langle R_{1,2}\rangle_s=0$, Leschke, Manai Ruder and Warzel have proven 
\begin{equation}
\displaystyle \liminf_{N\to\infty}\mathbb E\langle R_{1,2}^2 \rangle_s \geq 
F(2 \beta b\tanh \beta b)-\frac{2\kappa}{\beta \sqrt{s}}
,\label{LMRW}
\end{equation} 
 in the model defined by the Hamiltonian (\ref{hamils}) for $q=0$ \cite{W}, where
the function $F$ is defined by (\ref{FBF}). 
This inequality and the identity (\ref{RSparaeq}) imply the following theorem.
{\theorem \label{T3.4} (Non-exactness of the paramagnetic RS solution)\\
 If $\beta$ and $b$ satisfy %$\sqrt s \in [0,1]$ satisfies\begin{equation}
%\sqrt{s} > 
$\displaystyle\beta F(2 \beta b\tanh \beta b)> 2\kappa,$
%, \end{equation}
then  the inequality (\ref{RSpara}) becomes strict
$$\varphi_\infty(1) < \liminf_{N\to \infty}\Phi(\beta,b,0).$$
Proof.} Define $\displaystyle \sqrt{s_0}:= \frac{2 \kappa}{\beta F(2 \beta b \tanh \beta b)}$. 
Then, the assumption $s_0 <1$ and inequality (\ref{LMRW}) enables us to evaluate the deviation
\begin{eqnarray*}
&&\liminf_{N\to\infty}[\Phi(\beta, b, 0)-\varphi_N(1)] \geq
\liminf_{N\to\infty} \int_{s_0}^1ds \mathbb E \langle R_{1,2}^2\rangle_s \\
&&\geq\int_{s_0}^1ds \Big[F(2 \beta b\tanh \beta b)-\frac{2\kappa}{\beta \sqrt{s}}\Big]
= F(2 \beta b \tanh \beta b)(1-\sqrt{s_0})^2>0.
%\frac{[\beta F(2 \beta b \tanh \beta b)-2 \kappa]^2}{\beta^2 F(2 \beta b \tanh \beta b)}.
\end{eqnarray*}
This completes the proof. $\Box$\\

In this case, the bound $\Phi(\beta,b,0)$ becomes an approximate solution of $\varphi_\infty(1)$, and 
a better one may be given by a spin glass RS solution
$\Phi(\beta,b,q)$ with the minimizer $q>0$. Either spin glass RS or RSB phase is possible in this region of coupling constants, since there is no ferromagnetic long-range order in this model \cite{IISS}. 
After next section, we show in a different way that the spin glass RS solution
is not exact
$$
\displaystyle  \varphi_\infty(1) <\lim_{N\to\infty}\min_{q\in[0,1]}\Phi(\beta,b,q),
$$
like the paramagnetic RS one, and therefore the identity (\ref{R-q}) does not hold either. 
\\

 In the classical limit $b\to0$, the bound (\ref{Phi}) becomes
 \begin{equation}
\varphi_N(1) \leq  \mathbb E \log 2 \cosh \beta \sqrt{q}z
+ \frac{\beta^2}{4} (1-q)^2,
\end{equation}
which is identical to the RS solution in the SK model.
The equation
(\ref{Phi_q})  becomes 
 \begin{equation}
q= \mathbb E \tanh^2 \beta \sqrt{q}z . \label{classical}
\end{equation}
This has a solution $q=0$.
 In the classical case $b=0$,
 it was conjectured  that the replica symmetry is preserved  with 
 $$
 \lim_{N\to \infty}\mathbb E \langle (R_{1,2}-q)^2\rangle_1 =0, 
$$  and the SK solution of the free energy density is exact
 for
$$
\mathbb E \frac{\beta^2}{\cosh^4 \beta \sqrt{q}z } \leq 1,
$$
whose boundary is called  the AT line \cite{AT,Tn}. This condition becomes $\beta \leq 1$
for $q=0$.
Recently, Chen has proven rigorously that the SK solution is exact  in the classical model \cite{WK-C}.  

\section{1RSB bound on the free energy density}
Guerra obtained the RSB bound in the SK model in the square root interpolation \cite{G2}. 
This bound can find the AT line \cite{AT,T,Tn}.
Here, we extend this method  to the transverse field SK model and
demonstrate that a 1RSB solution gives better bound on the free energy density than the RS one (\ref{RS}).  
Assume the following square root interpolation of Hamiltonian with a parameter $s \in [0,1]$ between the transverse field Sherrington-Kirkpatrick model and
an independent spin model
\begin{equation}
H(s, \sigma, g,z,z^1):=-\sqrt{\frac{s}{N}} \sum_{1\leq i<j\leq N} g_{i,j}  \sigma_i^z \sigma_j^z
-\sqrt{1-s}\sum_{j=1}^N (\sqrt{q_1}  z_j + \sqrt{q_2-q_1}z_j^1)\sigma_j^z
-\sum_{j=1}^N b \sigma_j^x, \label{1RSBhamil}
\end{equation}
where variational parameters $q_1,q_2$ satisfy $0 \leq q_1 \leq q_2\leq 1$ and $z_j, z_j^1$ are i.i.d standard Gaussian random variables.
This interpolated  Hamiltonian for $b=0$ is identical to that in the SK model given in \cite{T}.  
Define a partition function 
\begin{equation}
Z(s):= {\rm Tr} e^{-\beta H(s, \sigma, g,z,z^1)}.
\end{equation}
Define an interpolation for a free energy density with another variational parameter $m \in [0,1]$
\begin{equation}
\psi_N(s):= \frac{1}{Nm}\mathbb E \log \mathbb E_1 Z(s)^m,
\end{equation}
where $\mathbb E_1$ denotes the expectation only over $(z^1_i)_{1\leq i\leq N}$ and 
 $\mathbb E$ denotes the expectation over all random variables.  
 Note that this function for $s=1$ is identical to the function (\ref{varphi1}) 
 \begin{equation}
 \psi_N(1) = \varphi_N(1),
 \end{equation}
 and $-\psi_N(1)/\beta$ is the free energy density of the transverse field SK model. 
The derivative of $\psi_N(s)$ is 
\begin{equation}
\psi_N'(s) = -\frac{\beta}{N} \mathbb E \frac{1}{\mathbb E_1 Z(s)^m}\mathbb E_1 Z(s)^m \langle \frac{\partial}{\partial s} H(s,\sigma,g,z,z^1) \rangle_s
=\rm I +II +III,
\end{equation}
where three terms are defined by
\begin{eqnarray}
{\rm I}&:=&\frac{\beta}{N} \mathbb E \frac{1}{\mathbb E_1 Z(s)^m} \mathbb E_1 Z(s)^m 
 \frac{1}{2\sqrt{sN}} \sum_{1\leq i<j\leq N} g_{i,j} \langle \sigma_i^z \sigma_j^z\rangle_s, \label{I}\\
{\rm II}&:=&-\frac{\beta}{N} \mathbb E \frac{1}{\mathbb E_1 Z(s)^m} \mathbb E_1 Z(s)^m 
\frac{1}{2\sqrt{1-s}}\sum_{j=1}^N \sqrt{q_1}  z_j \langle \sigma_j^z\rangle_s, \label{II}\\
{\rm III}&:=&-\frac{\beta}{N} \mathbb E \frac{1}{\mathbb E_1 Z(s)^m} \mathbb E_1 Z(s)^m 
\frac{1}{2\sqrt{1-s}}\sum_{j=1}^N \sqrt{q_2-q_1}  z_j^1 \langle \sigma_j^z\rangle_s. \label{III}
\end{eqnarray}
Integration by parts and $(\sigma_i^z \sigma_j^z,\sigma_i^z \sigma_j^z)_s\leq 1$ for the first term (\ref{I}) imply
\begin{eqnarray}
{\rm I}&=& \frac{\beta}{2N^\frac{3}{2}\sqrt{s}} \sum_{1\leq i<j\leq N}
\mathbb E \frac{\partial}{\partial g_{i,j}} \frac{1}{\mathbb E_1 Z(s)^m} \mathbb E_1 Z(s)^m 
 \langle \sigma_i^z \sigma_j^z\rangle_s \nonumber \\
 &=&\frac{\beta^2}{2N^2} \sum_{1\leq i<j\leq N}
\mathbb E\Big[-m \Big( \frac{1}{\mathbb E_1 Z(s)^m} \mathbb E_1 Z(s)^m 
 \langle \sigma_i^z \sigma_j^z\rangle_s\Big)^2\nonumber \\
 &+& \frac{m-1}{\mathbb E_1 Z(s)^m} \mathbb E_1 Z(s)^m 
 \langle \sigma_i^z \sigma_j^z\rangle_s^2 + \frac{1}{\mathbb E_1 Z(s)^m} \mathbb E_1 Z(s)^m (\sigma_i^z \sigma_j^z,\sigma_i^z \sigma_j^z)_s\Big]
 \nonumber \\
 &=&\frac{\beta^2(1-N)}{4N}
\mathbb E\Big[m \Big( \frac{1}{\mathbb E_1 Z(s)^m} \mathbb E_1 Z(s)^m 
 \langle \sigma_i^z \sigma_j^z\rangle_s\Big)^2\nonumber \\
 &+& \frac{1-m}{\mathbb E_1 Z(s)^m} \mathbb E_1 Z(s)^m  \langle \sigma_i^z \sigma_j^z\rangle_s^2 %+ 1
- \frac{1}{\mathbb E_1 Z(s)^m} \mathbb E_1 Z(s)^m (\sigma_i^z \sigma_j^z,\sigma_i^z \sigma_j^z)_s
 \Big].
\end{eqnarray}
Integration by parts and $(\sigma_j^z,\sigma_j^z)_s \geq \tanh \beta b/(\beta b)$ by
the Falk-Bruch inequality (\ref{FB}) 
for the second term (\ref{II}) imply
\begin{eqnarray}
{\rm II}&=&-\frac{\beta}{2N\sqrt{1-s}} \sum_{j=1}^N \mathbb E \frac{\partial}{\partial z_j}\frac{1}{\mathbb E_1 Z(s)^m} \mathbb E_1 Z(s)^m 
\sqrt{q_1}  \langle \sigma_j^z\rangle_s \nonumber \\
&=&\frac{\beta^2q_1}{2} \mathbb E \Big[ m\Big( \frac{1}{\mathbb E_1 Z(s)^m} \mathbb E_1 Z(s)^m 
 \langle \sigma_j^z\rangle_s \Big)^2+ \frac{1-m}{\mathbb E_1 Z(s)^m} \mathbb E_1 Z(s)^m 
 \langle \sigma_j^z\rangle_s^2\nonumber \\ &-& 
 \frac{1}{\mathbb E_1 Z(s)^m} \mathbb E_1 Z(s)^m 
 ( \sigma_j^z, \sigma_j^z)_s\Big]. 
\end{eqnarray}
The third term (\ref{III}) can be evaluated in the same way
\begin{eqnarray}
{\rm III}&=&-\frac{\beta}{2N\sqrt{1-s}} \sum_{j=1}^N \mathbb E\frac{1}{\mathbb E_1 Z(s)^m}\mathbb E_1  \frac{\partial}{\partial z_j^1} Z(s)^m 
\sqrt{q_2-q_1}  \langle \sigma_j^z\rangle_s \nonumber \\
&=&\frac{\beta^2(q_2-q_1)}{2} \mathbb E \Big[  \frac{1-m}{\mathbb E_1 Z(s)^m} \mathbb E_1 Z(s)^m 
 \langle \sigma_j^z\rangle_s^2 -
 \frac{1}{\mathbb E_1 Z(s)^m} \mathbb E_1 Z(s)^m 
 ( \sigma_j^z, \sigma_j^z)_s\Big]. 
\end{eqnarray}

Therefore, $\psi_N'(s)$ is represented as
\begin{eqnarray}
\psi_N'(s)&=&{\rm I + II+III} \nonumber \\
 &=&-\frac{\beta^2}{4}\Big[
m\mathbb E \Big( \frac{1}{\mathbb E_1 Z(s)^m} \mathbb E_1 Z(s)^m 
 \langle \sigma_i^z \sigma_j^z\rangle_s\Big)^2
 +(1-m)
\mathbb E\frac{1}{\mathbb E_1 Z(s)^m} \mathbb E_1 Z(s)^m  \langle \sigma_i^z \sigma_j^z\rangle_s^2 \nonumber \\
&-&2mq \mathbb E \Big( \frac{1}{\mathbb E_1 Z(s)^m} \mathbb E_1 Z(s)^m 
 \langle  \sigma_j^z\rangle_s\Big)^2 -2(1-m)q_2 \mathbb E \frac{1}{\mathbb E_1 Z(s)^m} \mathbb E_1 Z(s)^m 
 \langle  \sigma_j^z\rangle_s^2 \nonumber \\
 &-&1+2q_2
+ \rho_1(s,m,q_1,q_2) \Big], \label{I+II+III}
 \end{eqnarray}
where a non-negative valued function $\rho_1(s,m,q_1,q_2)$ is defined  by
 \begin{equation}
 \rho_1(s,m,q_1,q_2):=\mathbb E \frac{1}{\mathbb E_1 Z(s)^m} \mathbb E_1 Z(s)^m[ 2q_2(\sigma_j^z,\sigma_j^z)_s-2q_2
 +\frac{N-1}{N}[ 1- (\sigma_i^z\sigma_j^z,\sigma_i^z\sigma_j^z)_s]].
 \label{rho1}
 \end{equation}
 Inequalities (\ref{D1}), (\ref{D2} )and (\ref{Fineq})
 give the following uniform lower and upper bounds independent of $(s, m, q)$
 \begin{equation}
 2q_2\Big(\frac{\tanh \beta b}{\beta b} -1\Big) \leq  \rho_1(s,m,q_1,q_2) 
 \leq\frac{N-1}{N}\Big( 1- \frac{1-e^{-2\beta b\tanh \beta b}}{2\beta b\tanh \beta b}\Big).  \label{lurho1}
 \end{equation}
 Next, we represent the above bound on $\psi_N'(s)$ in a replicated model. Define a two
 replicated Hamiltonian
 by
 \begin{equation}
 H_2(s, \sigma^1,\sigma^2):=  H(s,\sigma^1,g,z^1)+ H(s,\sigma^2, g,z^2),
 \end{equation}
 where the right hand side consists of the interpolated one step RSB Hamiltonian defined by (\ref{1RSBhamil})
 with i.i.d.  standard Gaussian random variables $(z^a_i)_{1\leq i\leq N, a=1,2}$.
  Note that the partition function of this replicated model is factorized into the original partition functions
 \begin{equation}
 Z_{2}(s):={\rm Tr} e^{-\beta H(s,\sigma^1,\sigma^2)} = {\rm Tr} e^{-\beta H(s,\sigma, g,z^1)} {\rm Tr} e^{-\beta H(s,\sigma,g,z^2 )}.
 \end{equation}
 The following expectation of the overlap operator defined by (\ref{R}) is represented in terms of expectation values of the original model
 \begin{equation}
 \mathbb E \frac{1}{\mathbb E_1 \mathbb E_2 Z_{2}(s)^m}\mathbb E_1 \mathbb E_2 Z_{2}(s)^m\langle R_{1,2} \rangle_{s,2}=
 \mathbb E\Big( \frac{1}{\mathbb E_1Z(s)^m}\mathbb E_1 Z(s)^m\langle \sigma_i^z \rangle_s \Big)^2,
 \end{equation}
 where $\mathbb E_a$ denotes the expectation value only over $(z_i^a)_{1\leq i\leq N, a=1,2}$ and the Gibbs expectation value of $f(\sigma^1,\sigma^2)$ is defined by
 $$
 \langle f(\sigma^1,\sigma^2) \rangle_{s,2}:=\frac{1}{Z_{2}(s)} {\rm Tr} f(\sigma^1,\sigma^2) e^{-\beta H(s,\sigma^1,\sigma^2)}.
 $$
 Note also
  \begin{equation}
 \mathbb E \frac{1}{\mathbb E_1 \mathbb E_2 Z_{2}(s)^m}\mathbb E_1 \mathbb E_2 Z_{2}(s)^m\langle R_{1,2} ^2\rangle_{s,2}=\frac{N-1}{N}
 \mathbb E\Big( \frac{1}{\mathbb E_1Z(s)^m}\mathbb E_1 Z(s)^m\langle \sigma_i^z \sigma_j^z \rangle_s \Big)^2
 +\frac{1}{N}.
 \end{equation}
 These identities give
 \begin{eqnarray}
 && \mathbb E \frac{1}{\mathbb E_1 \mathbb E_2 Z_{2}(s)^m}\mathbb E_1 \mathbb E_2 Z_{2}(s)^m\langle( R_{1,2}-q_1 )^2\rangle_{s,2}
 \nonumber \\
 &&=\frac{N-1}{N} 
 \mathbb E\Big( \frac{1}{\mathbb E_1Z(s)^m}\mathbb E_1 Z(s)^m\langle \sigma_i^z \sigma_j^z \rangle_s \Big)^2
  -2q_1  \mathbb E\Big( \frac{1}{\mathbb E_1Z(s)^m}\mathbb E_1 Z(s)^m\langle \sigma_i^z \rangle_s \Big)^2
  +q_1^2+\frac{1}{N}.
  \label{Rq}
 \end{eqnarray}
 If the delta function is defined by
 \begin{equation}
 \delta(z^1,z^2) := \prod_{i=1}^N \sqrt{2\pi} e^\frac{(z_i^1)^2}{2}\delta(z_i^1-z_i^2),
 \end{equation}
 then 
 \begin{eqnarray*}
&& \mathbb E \frac{1}{\mathbb E_1 \mathbb E_2 Z_{2}(s)^\frac{m}{2}\delta(z^1,z^2) }\mathbb E_1 \mathbb E_2 Z_{2}(s)^\frac{m}{2}\langle R_{1,2} \rangle_{s,2}\delta(z^1,z^2) =\mathbb E \frac{1}{\mathbb E_1Z(s)^m}\mathbb E_1 Z(s)^m\langle \sigma_i^z \rangle_s ^2,  \\
&& \mathbb E \frac{1}{\mathbb E_1 \mathbb E_2 Z_{2}(s)^\frac{m}{2}\delta(z^1,z^2) }\mathbb E_1 \mathbb E_2 Z_{2}(s)^\frac{m}{2}\langle R_{1,2} ^2\rangle_{s,2}\delta(z^1,z^2) = \frac{N-1}{N} \mathbb E \frac{1}{\mathbb E_1Z(s)^m}\mathbb E_1 Z(s)^m\langle \sigma_i^z \sigma_j^z\rangle_s ^2+\frac{1}{N}.
 \end{eqnarray*}
 These identities give
 \begin{eqnarray}
 && \mathbb E \frac{1}{\mathbb E_1 \mathbb E_2 Z_{2}(s)^\frac{m}{2}\delta(z^1,z^2)}\mathbb E_1 \mathbb E_2 Z_{2}(s)^\frac{m}{2}\langle( R_{1,2}-q_2 )^2\rangle_{s,2}\delta(z^1,z^2)
 \nonumber \\
 &&=\frac{N-1}{N}
 \mathbb E \frac{1}{\mathbb E_1Z(s)^m}\mathbb E_1 Z(s)^m\langle \sigma_i^z \sigma_j^z \rangle_s^2
  -2q_2  \mathbb E \frac{1}{\mathbb E_1Z(s)^m}\mathbb E_1 Z(s)^m\langle \sigma_i^z \rangle_s^2+{q_2}^2+\frac{1}{N}. \label{R-q_2}
 \end{eqnarray}
 Identities (\ref{I+II+III}), (\ref{Rq}) and  (\ref{R-q_2}) enable us to represent the upper bound on $\psi_N'(s)$ 
 \begin{eqnarray}
 \psi_N'(s) =&-& \frac{\beta^2}{4}\Big[ m\mathbb E \frac{1}{\mathbb E_1 \mathbb E_2 Z_{2}(s)^m}\mathbb E_1 \mathbb E_2 Z_{2}(s)^m\langle( R_{1,2}-q_1 )^2\rangle_{s,2} \nonumber \\
&+&  (1-m)\mathbb E \frac{1}{\mathbb E_1 \mathbb E_2 Z_{2}(s)^\frac{m}{2}\delta(z^1,z^2)}\mathbb E_1 \mathbb E_2 Z_{2}(s)^\frac{m}{2}\langle( R_{1,2}-q_2 )^2\rangle_{s,2}\delta(z^1,z^2) \Big]
\nonumber\\
&+& \frac{\beta^2}{4}[m (q_1^2 -{q_2}^2)+(1-q_2)^2  -\rho_1(s,m,q_1,q_2)]. \label{phi'RSB}
 \end{eqnarray}
  Since the first and second terms in (\ref{phi'RSB}) are non-positive, the $\psi_N(1)$ is bounded by 
 \begin{equation}
 \psi_N(1) \leq  \psi_N(0) +
 \frac{\beta^2}{4}\Big[m (q_1^2 -{q_2}^2)+(1-q_2)^2  -
\int_0^1\rho_1(s,m,q_1,q_2)ds\Big].
 \end{equation}
 The partition function for $s=0$ can be calculated easily
 \begin{equation}
 Z(0)={\rm Tr } \exp \beta \sum_{i=1}^N [\sqrt{q_1}z_i\sigma_i^z +\sqrt{q_2-q_1} z_i^1 \sigma_i^z+b\sigma_i^x ] 
 =[2 \cosh Y(z,z^1,q_1,q_2)]^N,
 \end{equation}
 where the above random variable is defined by
 \begin{equation}
 Y(z,z^1,q_1,q_2):=\beta \sqrt{(\sqrt{q_1}z+\sqrt{q_2-q_1}z^1 )^2+b^2}.
 \end{equation} 
 Note that $q_1=q_2=q$  implies the following relation to the random variable defined by (\ref{X})
 \begin{equation} Y(z,z^1,q,q)=X(z,q). \label{XY} 
 \end{equation}
  %\newpage
 Define 1RSB bound by the following function
 \begin{eqnarray}
 &&\Psi(\beta,b,m,q_1,q_2):=  \\
 &&\frac{1}{m}\mathbb E \log \mathbb E_1  [2 \cosh Y(z,z^1,q_1,q_2)]^m +
 \frac{\beta^2}{4}\Big[m (q_1^2 -{q_2} ^2)+(1-q_2)^2  -
\int_0^1\rho_1(s,m,q_1,q_2)ds\Big]. \nonumber
\end{eqnarray} 
 The following lemma represents  $\psi_N(1)$ in terms of 1RSB bound.
 
 {\lemma (Extended Guerra's identity for 1RSB bound)\\
  \label{L4.1}For any $(\beta, b, m, q_1,q_2) \in[0,\infty)^2\times [0,1]^3$ $\psi_N(1)$ has an upper bound 
 \begin{eqnarray}
\psi_N(1) &=&  \Psi(\beta,b,m,q_1,q_2) 
- \frac{\beta^2}{4}\int_0^1ds\Big[ m\mathbb E \frac{1}{\mathbb E_1 \mathbb E_2 Z_{2}(s)^m}\mathbb E_1 \mathbb E_2 Z_{2}(s)^m\langle( R_{1,2}-q_1 )^2\rangle_{s,2} \nonumber \\
&+&  (1-m)\mathbb E \frac{1}{\mathbb E_1 \mathbb E_2 Z_{2}(s)^\frac{m}{2}\delta(z^1,z^2)}\mathbb E_1 \mathbb E_2 Z_{2}(s)^\frac{m}{2}\langle( R_{1,2}-q_2 )^2\rangle_{s,2}\delta(z^1,z^2) \Big]. 
 \label{Psi}
 \end{eqnarray} }
 
 Obviously, $\Psi(\beta,b,m,q_1,q_2)$   for any $(m,q_1,q_2)$  gives an upper bound on $\psi_N(1)$. 
  The inequalities (\ref{lurho1}) give the following lemma.
 
 {\lemma \label{L4.2} Lower and upper bounds on $ \Psi(\beta,b,m,q_1,q_2)$ are given by
 \begin{equation}
 \Psi_L(\beta,b,m,q_1,q_2) \leq \Psi(\beta,b,m,q_1,q_2) \leq \Psi_U(\beta,b,m,q_1,q_2),
 \end{equation}
 where above functions are defined by
 \begin{eqnarray}
  \Psi_L(\beta,b,m,q_1,q_2) 
  &:=& 
 \frac{1}{m} \mathbb E \log \mathbb E_1\cosh^m Y(z,z^1,q_1,q_2) \nonumber \\
  &+&\frac{\beta^2}{4}\Big[m (q_1^2 -{q_2}^2)+(1-q_2)^2 -\Big(1- \frac{1-e^{-2\beta b\tanh \beta b}}{2\beta b\tanh \beta b}\Big)\Big], \\ 
  \Psi_U(\beta,b,m,q_1,q_2)
  &:=& 
 \frac{1}{m} \mathbb E \log \mathbb E_1\cosh^m Y(z,z^1,q_1,q_2)   \nonumber \\
&+&\frac{\beta^2}{4}\Big[m (q_1^2 -{q_2}^2)+(1-q_2)^2+2q_2\Big(1 -\frac{\tanh \beta b}{ \beta b}\Big)\Big].
  \label{PsiU} 
\end{eqnarray}}
The identity (\ref{XY}) implies
that the 1RSB bound is identical to the RS bound defined by (\ref{Phi}) for  $q_1=q_2=q \in [0,1]$ for any $m\in [0,1]$, 
 \begin{equation}\Phi(\beta,b,q)=\Psi(\beta,b,m,q,q). \label{RS=1RSB}
 \end{equation}
 Define the 1RSB solution by
 \begin{equation}
 \min_{0 \leq m \leq 1,0\leq q_1\leq q_2\leq 1} \Psi(\beta,b,m,q_1,q_2). \label{1RSB}
 \end{equation}
 
 \section{AT-type instability}
 
 The lower bound (\ref{lPhi}) in Lemma \ref{L3.2} and the upper bound (\ref{PsiU}) in Lemma \ref{L4.2}
 enable us to obtain the following  theorem. 

{\theorem \label{MT} 
For any $(\beta,b,q,m,q_1,q_2) \in [0,\infty)^2\times [0,1]^4$ with $q_1\leq q_2$, the difference between the RS and the 1RSB variational solutions  has a lower bound 
\begin{eqnarray}
 \min_{q\in[0,1]}\Phi(\beta,b,q)- \Psi(\beta,b,m,q_1,q_2)
    \geq 
    \min_{q\in [0,1]} \Theta(\beta,b,q,m,q_1,q_2),
   \end{eqnarray}
   where the function in the right hand side is defined by
   \begin{eqnarray}
  \Theta(\beta,b,q,m,q_1,q_2)&:=&\Phi_L(\beta,b,q)- \Psi_U(\beta,b,m,q_1,q_2) .
 \end{eqnarray}}  
The RS solution  cannot be the exact solution,
if the 1RSB solution (\ref{1RSB}) gives better bound for $\psi_N(1)=\varphi_N(1)$ than the RS one (\ref{RS}). This corresponds to the AT-type instability.
Let us show this instability in the RS solution on the basis of the bound given by Theorem \ref{MT}.
For some $(m,q_1,q_2)\in [0,1]^3$ satisfying $q_1\leq q_2$,  the condition 
\begin{equation}
\min_{q \in [0,1]}
\Theta(\beta,b,q,m,q_1,q_2) > 0, 
\end{equation}
is sufficient for
the AT-type instability in the RS solution (\ref{RS}).  
Since the function $\Theta(\beta,b,q,m,q_1,q_2)$ is represented 
in terms of physical quantities of disordered single spin systems, its numerical calculation can be done easily. Numerical calculations by Mathematica for $\Theta(\beta,b,q,m,q_1,q_2)$ with its minimizer $q\in [0,1]$
at several points $(\beta,b)\in[0,\infty)^2$ are obtained as follows:
\begin{eqnarray}
\hspace{-1cm}&&\Theta(1/0.10,10^{-3},0.92,0.70,0.88,0.99)=3.60 \times 10^{-2},\label{data1} \\
\hspace{-1cm}&&\Theta(1/0.30,10^{-3},0.73,0.76,0.71,0.91)=4.64 \times 10^{-3},\\
 \hspace{-1cm}&&\Theta(1/0.50,10^{-3},0.53,0.78,0.51,0.64)=4.81 \times 10^{-4},\\
  \hspace{-1cm}&&\Theta(1/0.70, 10^{-3},0.32,0.90,0.31,0.38)=1.44 \times 10^{-5}, \\
   \hspace{-1cm}&&\Theta(1/0.90, 10^{-3},0.12,0.99,0.10,0.22)=1.50 \times 10^{-5}.\label{data5}
\end{eqnarray}
Therefore, a computer-assisted proof by simple calculations
shows that the AT-type instability exists in the RS solution for the transverse field SK model.

\section{Discussions}
In the present paper, the square root interpolation method developed by Guerra and Talagrand has been extended to a mean field quantum spin glass model. We have studied the transverse field Sherrington-Kirkpatrick (SK) model with the $\mathbb Z_2$-symmetry.   
First, we obtain the replica-symmetric (RS) bound $\Phi(\beta, b,q)$ for the logarithm of partition function per spin, where $\beta >0$, $b >0$ and $q\in [0,1]$ are inverse temperature, strength of the transverse field and a variational parameter, respectively. 
Theorem \ref{T3.3} shows 
that the RS bound $\Phi(\beta, b, 0)$ is the exact solution, if the replica-symmetry and the $\mathbb Z_2$-symmetry are unbroken in the paramagnetic phase. 
On the other hand,  
Theorem \ref{T3.4} indicates that 
this paramagnetic RS solution cannot be exact, if the variance of overlap $R_{1,2}$
 does not vanish in sufficiently low temperature and sufficiently weak transverse field \cite{W}.  
Next, we study whether the spin glass RS solution $\Phi(\beta,b,q)$ with a positive minimizer $q$  
can be exact in this low temperature region.  
We obtain also one step replica-symmetry breaking (1RSB) bound $\Psi(\beta, b,m,q_1,q_2)$ with variational parameters $(m,q_1,q_2)\in [0,1]^3$ satisfying $q_1\leq q_2$. Note that $\Psi(\beta,b,m,q,q)=\Phi(\beta,b,q)$ for any $m\in[0,1]$. Using the Falk-Bruch inequality \cite{FB},
we obtain Theorem \ref{MT}, which gives a bound on the difference $\min_q\Phi(\beta,b,q)-\Psi(\beta,b,m,q_1,q_2)$ in terms of  disordered single spin systems. On the basis of Theorem \ref{MT}, simple numerical calculations for disordered single spin systems indicate 
that the 1RSB bound gives better bound than the RS bound at several points.
These show the existence of the de Almeida-Thouless(AT)-type instability in the RS solution.  
 Although surely confirmed unstable region is quite narrow in the coupling constant space,
our result is consistent with recently obtained results including numerical simulations \cite{KZL,W,MRC,SGSDC,Y}.

 In the transverse field SK model under an applied
 $\mathbb Z_2$-symmetry breaking longitudinal field,  
there is no proof that the RS solution $\min_q\Phi(\beta, b, q)$ is exact even in the high temperature region, since $\mathbb E \langle R_{1,2} \rangle_s$ may depend on $s \in [0,1]$.  
In the low temperature region of this model, however, it can be confirmed numerically also that the 1RSB solution gives better bound than the RS solution 
in sufficiently weak longitudinal and transverse fields, as in the classical SK model. 
Then, the RS solution cannot be exact either under the applied longitudinal field.
It turns out that the existence of the AT-type instability is not sensitive against the application of any weak longitudinal field.  

It should be studied still whether the infinite RSB ($\infty$RSB) occurs
%the $k$-step RSB ($k$RSB) bound for $k>1$ gives better bound than the 1RSB solution 
in the transverse field SK model. Also 2RSB bound is confirmed numerically to be a better bound than 1RSB solution. Since a $k$RSB bound gives better bound than the $(k-1)$RSB solution in the classical 
SK model, there exists $b_0>0$, such that  a $k$RSB bound gives a better bound 
than the $(k-1)$RSB solution for any $b\leq b_0$ also in the transverse field SK model \cite{IFS}. 
Therefore, the $\infty$RSB solution is predicted to be exact in the transverse field SK model, as in the classical SK model. 
\\

\paragraph*{Acknowledgments}
It is pleasure to thank K. Hukushima, H. Leschke and M. Yamaguchi for enlightening discussions.
 C.I. is supported by JSPS (21K03393).
%I am grateful to M. Aoyagi for discussions in early stage of this work. 

\paragraph*{Conflict of interest statement}
The authors declare no conflicts of interest.

\paragraph*{Data availability statement}
The authors declare that the data (\ref{data1})-(\ref{data5}) in this study are openly available.
There are no other data% supplementary information files
.
%Data sharing not applicable to this article as no datasets were generated or analyzed
%during the current study and article describes entirely theoretical research.

%\newpage


\begin{thebibliography}{13}

\bibitem{AB}  Adhikaria, A., Brennecke, C., :{Free-energy of the quantum Sherrington-Kirkpatrick spin-glass model with transverse field} J. Math. Phys. 61, 083302,1-16 (2020).


\bibitem{AT} de Almeida, J. R. L., Thouless, D. J.  :{Stability of the Sherrington-Kirkpatrick solution of spin glass model}. 
J. Phys. A :Math.Gen. {\bf 11}, 983-990 (1978).



\bibitem{BT} Brankov, J.G.,  Tonchev,  N.S: {Generalized inequalities for the Bogoliubov-Duhamelinner product with applications in the ApproximatingHamiltonian Method}. Cond. Matt. Phys.{\bf 14},13003 (2011).



\bibitem{WK-C} Chen, W.-K., :{On the Almeida-Thouless transition line in the Sherrington-Kirkpatrick model with centered Gaussian external field.} 
 Electron. Commun. Probab. {\bf 26}, 
 65, 1-9 (2021).



\bibitem{Cr} Crawford, N. :{Thermodynamics and universality for mean field quantum spin glasses.} Commun. Math. Phys.\textbf{274}, 821-839(2007) 

\bibitem{DLS} Dyson, F. J.,  Lieb, E. H.,  Simon, B. :{Phase transitions in quantum spin systems with
isotropic and nonisotropic interactions} . J. Stat. Phys. {bf 18}, 335-383 (1978). 

\bibitem{FB} Falk, H., Bruch, L. W. :{Susceptibility and fluctuation}. Phys. Rev. {\bf 180}, 442-444 (1969).

 
\bibitem{G1} Guerra, F.: {Sum rules for the free energy in the mean field spin glass model}. Fields Inst. Commun. {\bf 30}, 161 (2001).

 
\bibitem{G2} Guerra, F.:{ Replica broken bounds in the mean field spin glass theory}. Commun. Math. Phys. {\bf 233},1-12,(2003).


\bibitem{IFS} Itoi, C., Fujiwara, K., Sakamoto, Y.:{Parisi-type formula in the transverse field 
Sherrington-Kirkpatrick model} in preparation.

\bibitem{IISS} Itoi, C., Ishimori, H., Sato, K., Sakamoto, Y. 
:{Universality of replica-symmetry breaking 
in the transverse field Sherrington-
Kirkpatrick model.}
J. Stat. Phys. {\textbf 190} 65, 1-9 (2023). 

\bibitem{IMT} Itoi, C., Mukaida, H., Tasaki, H. :{Griffiths-type theorem short-Rrange spin glass models.}
J. Stat. Phys. {\textbf 191} 28, 1-30 (2024). 
 
\bibitem{IS} Itoi, C., Sakamoto, Y. :{Boundedness of susceptibility in spin glass transition of transverse field mixed $p$-spin glass models}
JPSJ,{\bf 92}, 074001,1-11(2023)


 
  \bibitem{KZL} Kiss, A., Zarand, G., Lovas, I. :{Complete replica solution for the transverse field Sherrington-Kirkpatrick spin glass model with continuum-time quantum Monte Carlo method}. Phys. Rev. B {\bf 109}, 024431(1-20) (2024).


\bibitem{W} Leschke, H., Manai, C., Ruder, R., Warzel, S. : {Existence of RSB in quantum glasses}. Phis. Rev. Lett. {\bf 127}, 207204,1-6 (2021).      

\bibitem{LRRS} Leschke, H., Rothlauf, S.,  Ruder, R.,  Spitzer, W. :{The free energy of a quantum Sherrington-Kirkpatrick spin-glass model  for weak disorder}, J. Stat. phys. {\textbf 182} 55,1-41 (2021) 



\bibitem{MRC} Mukherjee, S.,  A. Rajak, A.,   Chakrabarti, B. K. :{Possible ergodic-nonergodic regions in the quantum Sherrington-Kirkpatrick spin glass model and quantum annealing}. Phys.Rev. E {\bf 97}, 022146,1-6(2018).




\bibitem{R} Roepstorff, G. :{A stronger version of Bogoliubov's  inequalities}.    Commun. Math. Phys.\textbf{53}, 143-150 (1977).



\bibitem{S} Shastry, B. S., :{Bounds for correlation functions of the Heisenberg antiferromagnet}.  J. Phys. A: Math. Gen. \textbf{25}L249-L253(1992).



\bibitem{SGSDC}Schindler, P. M., Guaita, T., Shi, T., Demler, E., Cirac, J. I.
 :{Variational Ansatz for the Ground State of the Quantum Sherrington-Kirkpatrick Model}.
 Phys. Rev. Lett. {\bf 129} 220401,1-6 (2022).

\bibitem{SK} Sherrington, S., Kirkpatrick, S. :{ Solvable model of spin glass}. Phys. Rev. Lett. \textbf{ 35},  1792-1796, (1975). 
 
  
 

\bibitem{T} Talagrand, M. :{ Mean field models for spin glasses I, II}.
Ergebnisse der Mathematik und ihrer Grenzgebiete. 3. Folge A Series of Modern Surveys in Mathematics, Vol. 54, 55. 
Springer-Verlag, Berlin (2011).
	

\bibitem{Tn}Toninelli, F., :{About the Almeida-Thouless transition line in the Sherrington-Kirkpatrick mean field spin glass model.} Euro.Phys. Lett. {\bf 60},5,764-767 (2002)  

\bibitem{Y}Young, A. P :{Stability of the quantum Sherrington-Kirkpatrick spin glass model.} Phys. Rev. E {\bf 96} 032112, 1-6(2018).



\end{thebibliography}
\end{document}